\documentstyle[fortschritte,amssymb,amsmath,epsf]{article}
\begin{document} 
\title{Entanglement at distance: qubits versus continuous
variables.}
\author{G. M. D'Ariano and M. F. Sacchi\\
Quantum Optics \& Information Group \\
Dipartimento di Fisica `A. Volta', Universit\`a di Pavia 
and INFM,\\  via A. Bassi 6, I-27100 Pavia, Italy}
\maketitle
\begin{abstract}
We consider the problem of obtaining maximally entangled photon states
at distance in the presence of loss. We compare the efficiency of two
different schemes in establishing $N$ shared ebits: 
i) $N$ single ebit states with the qubit encoded on polarization;   
ii) a single continuous variable entangled state (emode) assisted by
optimal local operation and classical communication (LOCC) protocol in
order to obtain a $2^N$-dimensional maximally entangled state, with
qubits encoded on the photon number.
\end{abstract}
\section{Introduction}

The production of maximally entangled state of photons at distance is
a key issue in communications of quantum information, for distributed
quantum computation \cite{pope}, quantum teleportation\cite{ben}, and
quantum cryptography\cite{ek}. Unfortunately, the detrimental effect
of losses is a serious problem for establishing entangled resources at
distance, since any kind of non-classical state is very sensitive to
the effect of loss.  

\par If one needs to teleport $N$ qubits from Alice to Bob, $N$ ebits
need to be shared between them, and for such purpose photons are the
only practically available carriers. One can use equivalently either
$N$ single ebit with the qubit encoded on polarization, or a single
continuous variable entangled state---''emode''-with qubits encoded
on the photon number. Parametric downconversion allows to create both
kinds of entangled states, ebits and emodes, in the low and high
gain regimes, respectively. 

\par In this paper we compare ebits and emodes in the presence of
loss. In contrast to the case of a single emode,  a scheme based on ebits
with the qubit encoded on the polarization of single photons has the
obvious advantage that the successful achievement of the ebit is
automatically checked by the presence of the photon itself at both
Alice and Bob sites, whereas for a single emode, this is not possible,
due its vacuum component. While there is no viable method for testing
the presence of the emode without destroying the entanglement, a scheme
for purification of emodes in the presence of loss has been proposed
in Ref. \cite{zoller}, however, with the disadvantage of achieving a
maximally entangled state of a random set of modes, hence with the
need of changing the encoding/decoding procedure each time, depending 
on which are the entangled modes. Therefore, in the presence of loss
only the ebit allows {\em knowingly} successful
teleportation/communication without increasing the complexity of the 
protocol versus $N$. Since emodes are more sensitive to loss for
increasingly large number of photons, a way out for implementing
emodes in the presence of moderate losses is to produce weakly
entangled modes, then performing a LOCC operation to enhance the
entanglement at the output of the lossy channel: this is the only
viable method for designing a protocol based on ebits with low
probability of failure. 

\par In this paper we will compare the efficiency of ebits and emodes
in establishing $N$ maximally entangled ebits, by considering a
protocol for emodes in which weakly entangled states are prepared
and an optimal LOCC operation is performed after the loss in order to obtain a
$2^N$ maximally entangled state. As we will see, the ebits are largely
superiors to emodes in all cases.

\section{The comparison}
Our task is to create $N$ ebits shared at distance between Alice and
Bob in the presence of loss.  $N$ ebits are represented by 
$N$ copies of a maximally entangled state belonging to the
tensor-product ${\mathbb C} ^2\otimes {\mathbb C} ^2$ of two-dimensional
Hilbert spaces, or equivalently by a single maximally entangled state
in ${\mathbb C}^{2^N}\otimes{\mathbb C}^{2^N}$. To achieve this task
we consider the use of $N$ ebits and the use of a single emode, with qubits
encoded on photon polarization and photon number, respectively. 
\par First we consider the use of $N$ single-photon states
\begin{eqnarray}
|\psi \rangle =\frac {1}{\sqrt 2}(|0 \rangle _a  |1 \rangle _b  +
|1 \rangle _b  |0 \rangle _b ) \;,\label{psi}
\end{eqnarray}
where the subscripts $a,b$ denote Alice and Bob sites.
\par The effect of loss on a single-mode state $\rho $ is described by
a completely positive map that can be written in the Kraus form \cite{pl}
\begin{eqnarray}
{\cal L}[\rho ]&\doteq& \sum_{n=0}^{\infty}
V_n \rho V^\dag_n\;,\end{eqnarray}
with
\begin{eqnarray}
V_n&=&\frac{(\eta^{-1}-1)^{n/2}}{\sqrt{n!}}a^n \eta ^{1/2 a^\dag
a}\;,\qquad 0\leq \eta \leq 1 \label{loss}
\end{eqnarray}
The physical parameter $\eta $ plays the role of the energy attenuation
factor, since on has $\mbox{Tr}[{\cal L}[\rho ]a^\dag a]=\eta \mbox{Tr}[\rho
a^\dag a]$. The smaller is the value of $\eta $, the larger is the effect
of the loss. More generally, $\eta $ gives the scaling factor of any normal-ordered 
operator function, namely
\begin{eqnarray}
{\cal L}^{\vee }\mbox{\bf :} f(a^{\dag} ,a) \mbox{\bf :} =\mbox{\bf :}
f(\eta ^{1/2}a^{\dag} ,\eta ^{1/2}a)\mbox{\bf :} \;,\label{ord}
\end{eqnarray}
where ${\cal L}^{\vee }$ denotes the dual map, which
is defined through the identity
\begin{eqnarray}
\mbox{Tr} \left[ {\cal L}^{\vee }[O] \rho \right]
=\mbox{Tr}\left[ O {\cal L}[\rho ]\right]  \; 
\end{eqnarray}
valid for any operator $O$. 

\par The typical best achievable value of the loss in a optical fibers
is of order $0.3\ dB/km$. Hence for two parties $10km$ far apart the loss is $3dB$,
corresponding to $\eta =1/2$.  \par Each mode is affected by the
effect of loss, hence from the maximally entangled state (\ref{psi})
one obtains the mixture
\begin{eqnarray}
{\cal L}_a\otimes {\cal L}_b[|\psi \rangle \langle \psi |]=\eta |\psi
\rangle \langle \psi |+(1-\eta ) |0\rangle _a {}_a\langle 0 |\otimes
|0\rangle _b {}_b\langle 0 | \;.\label{mix}
\end{eqnarray}
Therefore, the probability of sharing $N$ maximally entangled ebits in
the presence of loss is given by $p_b=\eta^N$.   

\par Now we consider the second scheme, based on a single emode (``twin-beam'') 
\begin{eqnarray}
|\chi (\lambda )\rangle =\sqrt{1- |\lambda |^2}\sum_{i=0}^\infty \lambda ^i |i
 \rangle _a |i \rangle _b\;,\qquad |\lambda |<1\;.\label{tb}
\end{eqnarray}
When producing the state (\ref{tb}) by parametric down-conversion, the
value of parameter $\lambda $ is related to the gain $G$ of the
optical amplifier as $G=(1-|\lambda | ^2)^{-1}$. Typically, one has
$|\lambda | =0.2 \div 0.75$ \cite{dema}. The state (\ref{tb}) is the
entangled resource for the continuous variable teleportation of
Ref. \cite{braun}.  In a way analogous to Eq. (\ref{mix}), the
twin-beam state that has suffered the effect of loss becomes a
mixture, and here we are interested only in the component that is
still a twin-beam, which is given by
\begin{eqnarray}
V_0\otimes V_0 |\chi (\lambda )\rangle\langle \chi (\lambda )| V^\dag
_0\otimes V^\dag _0= \frac {1- |\lambda |^2}{1-\eta |\lambda |^2}|\chi
(\eta ^{1/2}\lambda )\rangle\langle \chi (\eta ^{1/2}\lambda )|
\;.
\end{eqnarray}
We rewrite the state Eq. (\ref{tb}) damaged by the loss as follows
\begin{eqnarray}
{\cal L}_a\otimes {\cal L}_b[ |\chi (\lambda )\rangle\langle \chi
(\lambda )|]=q\,|\chi (\eta ^{1/2}\lambda )\rangle\langle \chi
(\eta ^{1/2}\lambda )| + \sigma 
\;,\label{}
\end{eqnarray}
with
\begin{eqnarray}
q =\frac {1- |\lambda |^2}{1-\eta |\lambda |^2} \;,\label{q}
\end{eqnarray}
and $\sigma $ is a positive operator with $\mbox{Tr}[\sigma ]=1-q$.
The value $q $ gives the probability that the twin-beam state
survives the loss, a part from the gain rescaling $\lambda
\rightarrow \eta ^{1/2}\lambda $.  \par Our task is now to achieve the
maximally entangled state
\begin{eqnarray}
|\phi \rangle =\frac {1}{\sqrt M}\sum_{i=0}^{M-1} |i \rangle _a |i
 \rangle _b\;.\label{m}
\end{eqnarray}
For this purpose we perform a LOCC transformation on the
state $|\chi(\eta ^{1/2}\lambda )\rangle $.  From Vidal's theorem
\cite{vidal}, we know that the maximal probability $p^*$ of obtaining
the state $|\phi \rangle $ from $|\chi \rangle $ by means of a LOCC
is given in terms of the Schmidt coefficients 
$\{\phi _i\}$ and $\{\chi _i\}$ of the states by
\begin{eqnarray}
p^*=\min _{i}\frac{\sum _{n=i} ^\infty \chi ^2_n} {\sum _{n=i} ^\infty
\phi ^2_n} \;.\label{}
\end{eqnarray}
In our case one has 
\begin{eqnarray}
p^*= \min _{i\in [0,M-1]}\frac{M (\eta |\lambda |^2)^{i}}{M-i} \leq M
(\eta |\lambda |^2)^{(M-1)}\equiv p' \;.
\end{eqnarray}
Moreover, one can easily show that for $|\lambda | \leq (\eta
e)^{-1/2}$ one has $p^*=p'$.  Hence, the overall probability $p_m$ of
obtaining the maximally entangled state $|\phi \rangle $ using a
twin-beam $|\chi (\lambda )\rangle $ in the presence of loss 
by means of an optimal LOCC transformation is given by
\begin{eqnarray}
p_C=q \,p^*=\frac {1- |\lambda |^2}{1-\eta |\lambda
|^2}\,p^*\;.\label{pc}
\end{eqnarray}
In order to compare $N$ ebits versus a single emode one takes $M=2^N$ for the
state (\ref{m}) and compares the probabilities $p_b$ (ebits) with $p_m$
(emodes).  Some numerical results are shown in Fig. 1, where the
probabilities $p_b$ (circles) and $p_m$ (triangles) are reported for
different values of $\eta$ and $\lambda $.
The comparison is dramatic: qubits are much more efficient than emodes
for any realistic value of the gain parameter $\lambda $ and loss $\eta $.
\begin{figure}[hbt]
\begin{center}\epsfxsize=.35 \hsize\leavevmode\epsffile{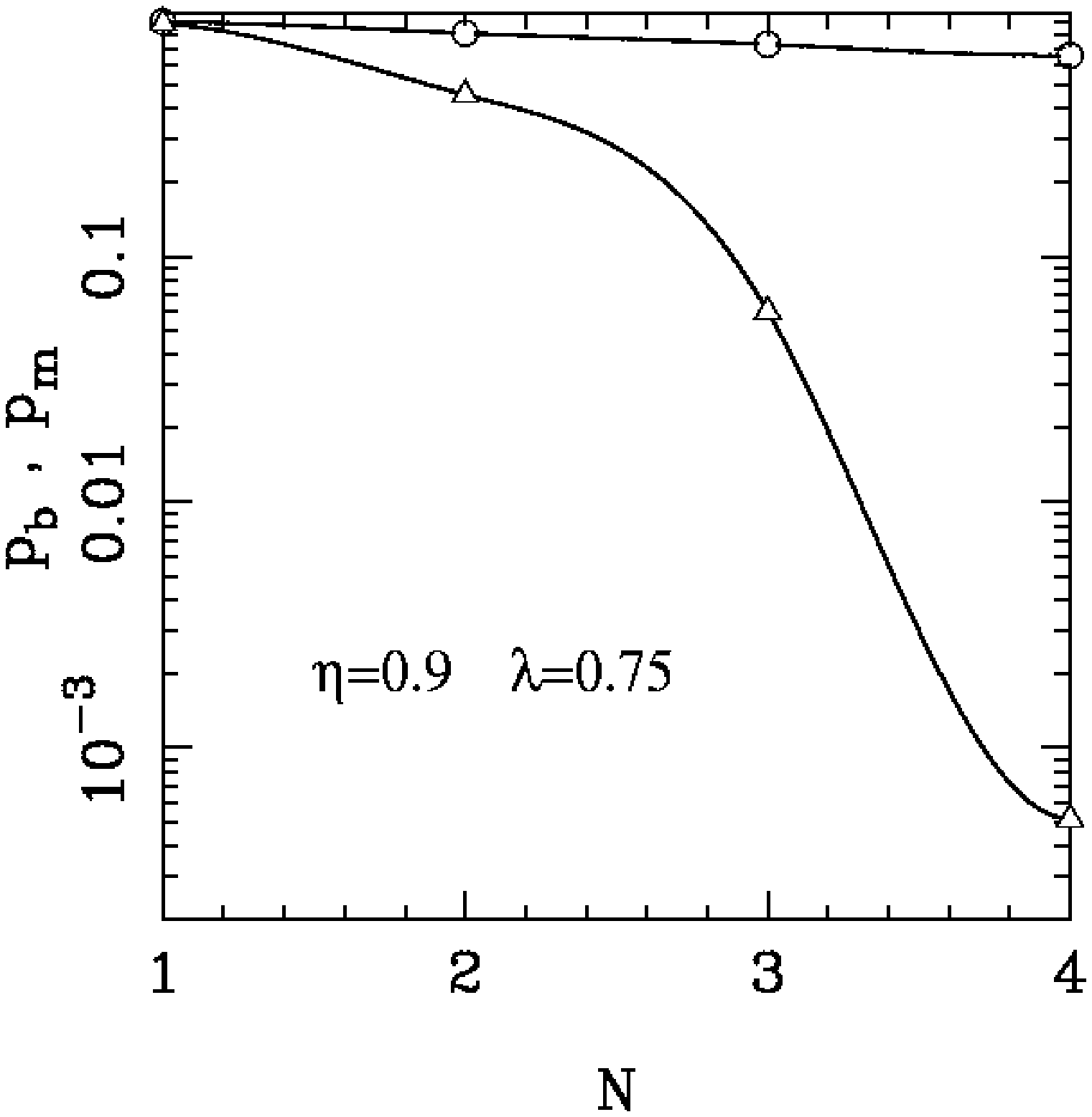}
\hspace{40pt}\epsfxsize=.35 \hsize\leavevmode\epsffile{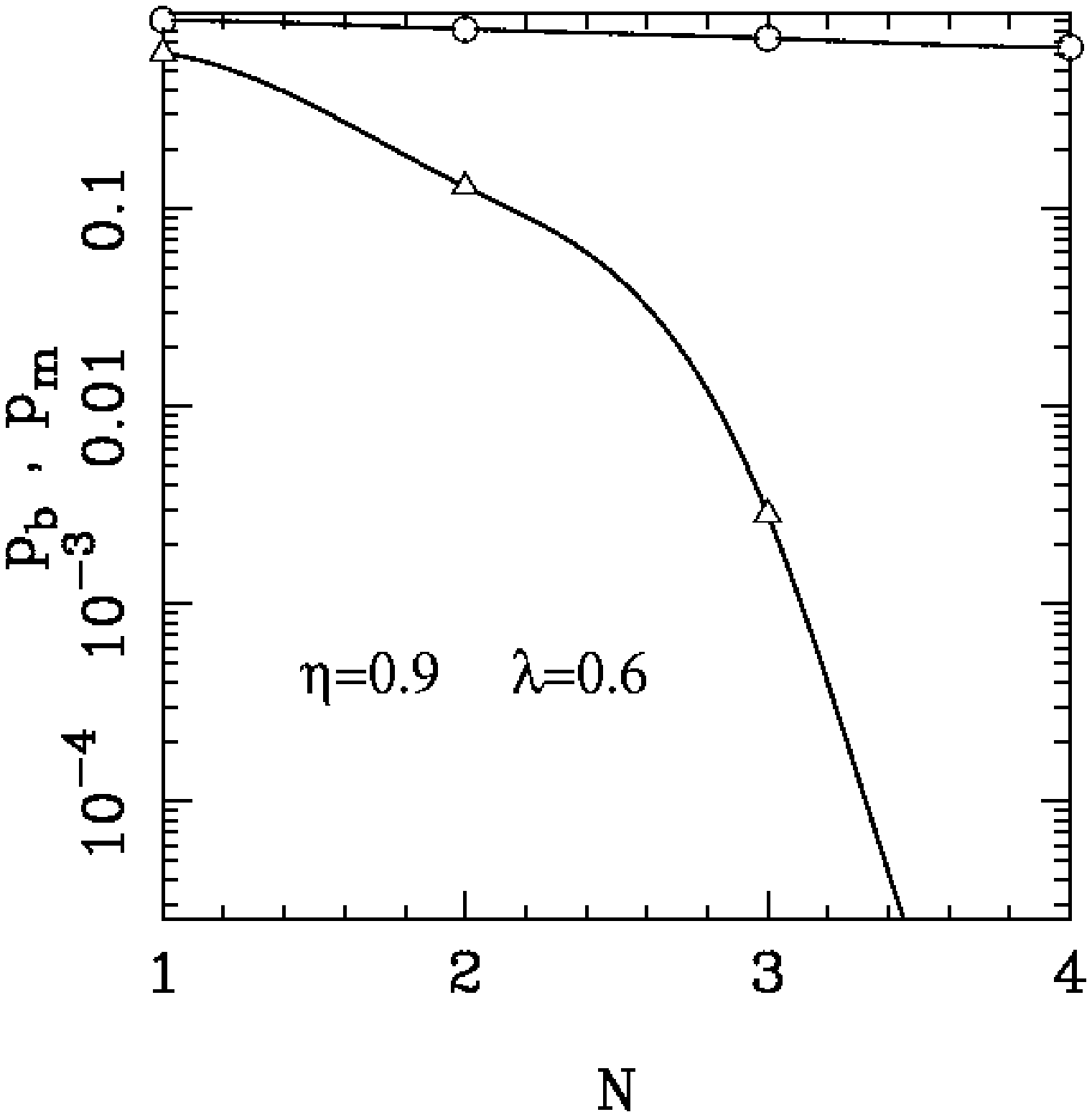}
\end{center}
\begin{center}\epsfxsize=.35 \hsize\leavevmode\epsffile{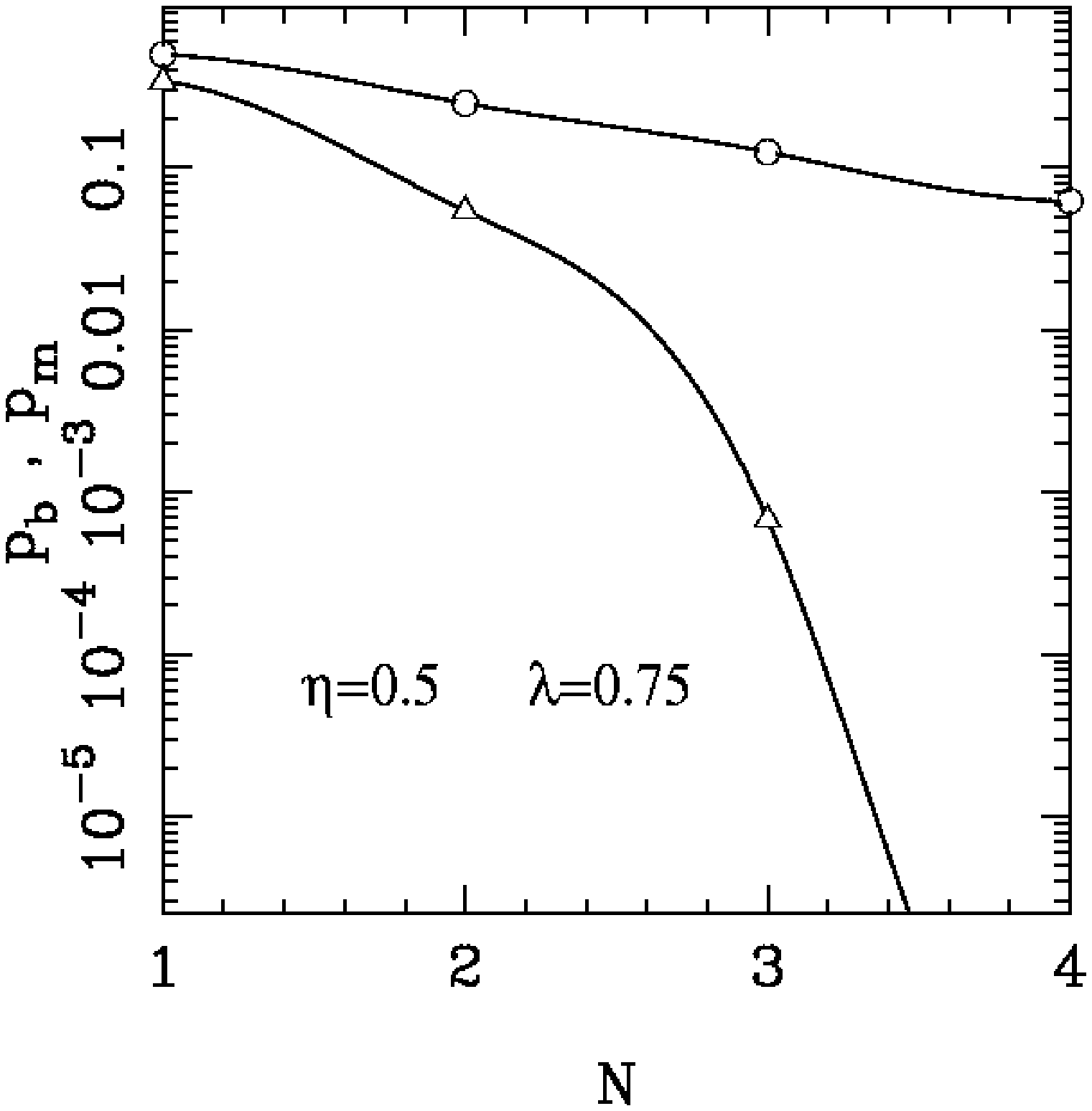}
\hspace{40pt}\epsfxsize=.35 \hsize\leavevmode\epsffile{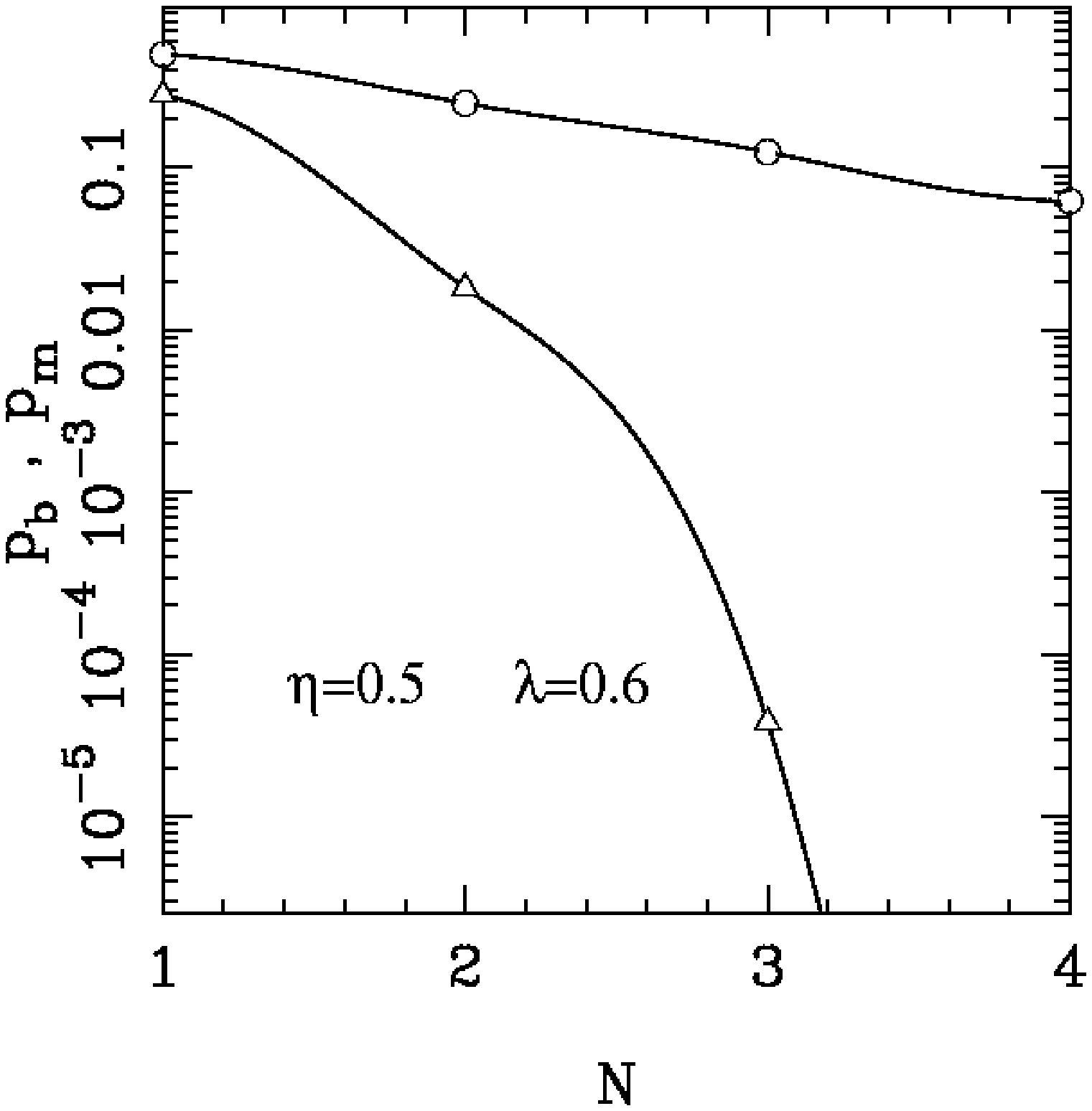}
\end{center}
\caption{Probability of successfully obtaining $N$ shared ebits
in the presence of loss using: i) (circles) $N$ single ebit states with the
qubit encoded on polarization;  ii) (triangles) a single continuous
variable entangled state (emode) assisted by optimal local operation
and classical communication (LOCC) protocol in order to obtain a 
$2^N$-dimensional maximally entangled state, with qubits encoded on
the photon number. We considered different values of the loss $\eta$
and of the gain parameter of emodes $|\lambda |$ (twin-beams in
Eq. (\ref{tb})).}\label{f:scheme}\end{figure}
\par The much greater efficiency of ebits versus emodes can be
inspected analytically as follows. We define the ratio $r\equiv p_b/p'$
and we have the chain of inequalities
\begin{eqnarray}
r=\frac {p_b}{p'}\leq \frac {p_b}{qp'}\leq \frac {p_b}{qp^*}=\frac
{p_b}{p_m}\;.\label{rat}
\end{eqnarray}
From Eq. (\ref{rat}) it is clear that $r >1$ implies $p_b>p_m$, namely
the scheme based on $N$ ebits is more efficient than that based on a
single emode. The ratio $r$ writes as 
\begin{eqnarray}
r=\left(\frac{\eta }{2}\right )^N \left( \frac{1}{\eta |\lambda |^2}\right
)^{2^N -1}=
\exp\{(1-2^N)\ln (\eta |\lambda |^2)  +N\ln (\eta /2)\}\;.
\label{ratio}
\end{eqnarray}
\par The expression of $r$ shows that for sufficiently large $N$ one
has $r>1$, and  the use of $N$ ebits becomes rapidly much more
efficient than using a single emode. In addition, we want to emphasize
again that in the presence of loss only the ebit allows {\em
knowingly} successful teleportation/communication without increasing
the complexity of the protocol versus $N$.
\section{Conclusions}
We have considered the problem of obtaining maximally entangled photon
states at distance in the presence of loss, and compared the
efficiency in establishing $N$ shared ebits by using $N$ single ebit
states---with the qubit encoded on polarization---and a single
continuous variable entangled state (which we called ``emode'') assisted by 
an optimal LOCC protocol in order to obtain a $2^N$-dimensional maximally
entangled state, with qubits encoded on the photon number. We have
shown the dramatic superiority of $N$ ebits versus a single emode,
besides the fact that only the ebit allows {\em knowingly} successful
teleportation/communication. We have not considered the possibility of
purification schemes for emodes. However, we emphasize again that the
only proposed scheme \cite{zoller} has the disadvantage of achieving a
maximally entangled state of a random set of modes, with the need of 
encoding/decoding procedures whose complexity is increasing versus $N$.
We conclude that a fruitful use of twin beams in quantum information
technology at distance completely relies on the possibility of
practical purification schemes, which need to be properly designed in a
way which is suitable to the particular entanglement-based protocol of
interest.
\section*{\protect{\noindent\normalsize\bf Acknowledgments}}
One of us (G. M. D. ) acknowledges stimulating discussions with
H. P. Yuen. This work has been founded by the EC program ATESIT, Contract
No. IST-2000-29681, and by DARPA Grant No. F30602-01-2- 0528.

\end{document}